\documentclass[aps,prl,preprint,superscriptaddress]{revtex4}

\begin{document}


\title{General condition of the Mott transition expressed with the density of state}


\author{Kazunori Shibata}
\email[]{ka-shibata@ppc.osaka-u.ac.jp}
\affiliation{Photon Pioneers Center, Osaka University, 
2-1 Yamada-oka, Suita Osaka 565-0871, Japan}

\author{Hiroaki Kakuma}
\email[]{d08001p@math.nagoya-u.ac.jp}
\affiliation{Graduate school of mathematics, Nagoya University,
Furo-cho, Chikusa-ku, Nagoya, 464-8602, Japan}

\author{Ryosuke Kodama}
\email[]{kodama@eie.eng.osaka-u.ac.jp}
\affiliation{Photon Pioneers Center, Osaka University, 
2-1 Yamada-oka, Suita Osaka 565-0871, Japan}
\affiliation{Graduate School of Engineering, Osaka University, 
2-1 Yamada-oka, Suita Osaka 565-0871, Japan}


\date{\today}

\begin{abstract}
We have investigated a contribution of the density of state (DOS) 
to the Mott metal-insulator transition (MIT). 
The MIT condition is given by the DOS, which usually requires the complicated fundamental equations. 
To simplify the process, we have derived a general solution of the MIT condition by 
an integral contains the DOS in the framework of coherent potential approximation. 
The formula is concise and presents the direct relationship between the MIT condition and the DOS. 
It is able to calculate the contribution quantitatively and 
to compare various DOSs analytically. 
\end{abstract}


\maketitle

The electron correlation brings about various phenomena \cite{2007Lohneysen,1991Sigrist}
which are beyond the one-electron approximation. 
One of the most wide-ranging and historical issues is 
the Mott metal-insulator transition (MIT) \cite{1968Mott}. 
In order to discuss the MIT, for example, coherent potential approximation (CPA) 
\cite{1963Hubbard, 1967Soven, 2002Bruno, 2004Kakehashi}, 
dynamical mean field theory (DMFT) \cite{1989Metzner, 1996Georges, 1998Imada, 2004Craco, 2006Lombardo} 
and their combination are used \cite{2001Laad}.
For both of the methods, the single electron density of state (DOS) is an important factor for the MIT. 
The DOS is composed through all the eigenvalues of a model Hamiltonian for a single electron \cite{1996Blawid}. 
In this manner, the DOS is inseparably related to the Hamiltonian. 
Therefore, the condition for the MIT takes place varies by the DOS. 
For example, it can be qualitatively expected a material becomes a metallic state easily 
if it has more eigenstates near the band edge rather than around the band center. 
However, it is impossible to conclude immediately how the MIT condition depends on the DOS 
because the fundamental equations appear in CPA 
or DMFT are complicated \cite{1996Georges, 1998Imada}. 
The numerical calculations, recent general approach, 
are also unable to give an answer for the question and the general dependence on the DOS. 
The numerical calculations are effective to treat the specified DOS. 
On the other hand, the analytical treatment is possible to handle the DOS in general form.

Here, we have proven that the MIT condition is simply given by the DOS in the framework of CPA.
We have derived the formula of the MIT condition for the symmetric DOS 
(e.g., it is valid for such as simple cubic and body centered cubic structures 
\cite{1989Metzner,1989Lohrmann}), 
in the case of paramagnetic state and zero temperature. 
The organization of this paper is as follows. 
First of all, we begin with a nondimensionalization of fundamental equations used in CPA. 
From the equations, we derive the necessary and sufficient conditions in order to constitute 
the energy band of electron. 
Using the conditions, we derive the MIT condition which is related to the DOS.

To describe the Mott MIT using CPA, 
electronic Green function and the self energy for spin $ \sigma $ is determined 
self-consistently through these two equations \cite{1973Yonezawa1, 1973Yonezawa2, 1996Blawid},
\begin{equation}\label{CPA01-01}
\Sigma(\omega)=\frac{nU}{1-(U-\Sigma(\omega))G(\omega)}, \ \ \ \ 
G(\omega)=\int \frac{\rho_0'(E)\text{d}E}
{\omega-E-\Sigma(\omega)+i\delta}.
\end{equation}
Here, $ \omega $ is energy of single electron, $ G(\omega) $ is the single electron Green function 
and $ \Sigma(\omega) $ is the self energy. $ U $ is Hubbard U that represent repulsive potential 
between two electrons on the same site.
$ n $ is average number of electron whose spin is $ -\sigma $. 
We treat $ n=1/2 $ for paramagnetic state.
$ \rho_0'(E) $ is normalized electronic DOS with no interaction between them.
After solving the simultaneous equations, we should consider for $ \delta \to +0 $.
We begin our analytical treatment with the nondimensionalization of these 
fundamental equations (\ref{CPA01-01}).
Generally, $ \rho_0'(E) \ge 0 $ at $ E_0-D \le E \le E_0+D $ and $ \rho_0'(E) = 0 $ elsewhere. 
$ D $ is half width of the band width $ W $, i.e., $ D=W/2 $. 
Introducing a dimensionless DOS $ \rho_0(x)=D\rho_0'(E) $ and $ x=(E-E_0)/D $, 
$ \int \rho_0'(E)\text{d}E=\int_{-1}^1\rho_0(x)\text{d}x=1 $ is obtained.
For other variables, we define dimensionless variables as 
$ u=U/D, \ \alpha=(\omega-E_0)/D-u/2, \ a_{\varepsilon}(\alpha)=\text{Re}\Sigma(\omega)/D-u/2, \ 
b_{\varepsilon}(\alpha)=\text{Im}\Sigma(\omega)/D $, and $ \varepsilon=\delta/D $.
The dimensionless energy of single electron is expressed by $ \alpha $. 
$ a_{\varepsilon}(\alpha) $ and $ b_{\varepsilon}(\alpha) $ 
correspond to the real and imaginary part of the self energy, respectively 
(we omit the argument $ \alpha $). 
Also, we express dimensionless Green function as $ DG=G_R+iG_I $ ($ G_R,G_I \in R $). 
For simplicity, we omit the subscript $ \varepsilon $ for $ G_R $ and $ G_I $. 
Introducing $ \xi_{\varepsilon}=\alpha-a_{\varepsilon} $, we can rewrite the Green function as 
\begin{equation}\label{CPA01-02}
G_R=\int^1_{-1} \frac{(\xi_{\varepsilon}-x)\rho_0(x)}
{(x-\xi_{\varepsilon})^2+(b_{\varepsilon}-\varepsilon)^2}\text{d}x,\ \ 
G_I=\int^1_{-1} \frac{(b_{\varepsilon}-\varepsilon)\rho_0(x)}
{(x-\xi_{\varepsilon})^2+(b_{\varepsilon}-\varepsilon)^2}\text{d}x.
\end{equation}
Thus, dimensionless fundamental equations are obtained as, 
\begin{equation}\label{CPA01-03}
G_Ra_{\varepsilon}^2+\left[
1-2b_{\varepsilon}G_I\right]a_{\varepsilon}
-\left(b_{\varepsilon}^2+\frac{u^2}{4}\right)G_R=0, \ \
G_Ia_{\varepsilon}^2+2b_{\varepsilon}G_Ra_{\varepsilon}
+b_{\varepsilon}-\left(b_{\varepsilon}^2+\frac{u^2}{4}\right)G_I=0.
\end{equation}
The limiting values of $ a_{\varepsilon} $, $ b_{\varepsilon} $, and 
$ \xi_{\varepsilon} $ as $ \varepsilon \to +0 $ will be denoted by 
$ a $, $ b $, and $ \xi $ respectively if they exist. We define ($ a,b,\xi $) as 
the solution of Eqs. (\ref{CPA01-03}). 

The electronic DOS with interaction between them is defined as 
$ \rho(\alpha)=-1/\pi\lim_{\varepsilon \to +0} G_I(\alpha) \ge 0 $\cite{1992Jarrell, 1998Bulla}. 
This definition requires $ b \le 0 $.
It is known if $ \rho(\alpha) > 0 $, there is an electron which possesses energy $ \alpha $.
This energy region forms an energy band. 
For example, if and only if $ u=0 $, $ \rho(\alpha)=\rho_0(\alpha) $ is obtained and 
there is a single band. 
In order to discuss the Mott MIT, we need to figure out the energy band for $ u>0 $. 
By the definition of $ \rho(\alpha) $, the solution of Eqs. (\ref{CPA01-03}) 
that satisfies $ \rho(\alpha) > 0 $ is 
confined to the next two cases: (i) $ b \neq 0 $ and 
(ii) $ b=0 $, $ |\xi|<1 $. 
In the latter case, we obtain 
\begin{equation}\label{CPA01-04}
\left(a^2-\frac{u^2}{4}\right)G_R+a=0, \ \
\left(a^2-\frac{u^2}{4}\right)G_I=0.
\end{equation}
$ G_I \neq 0 $ requires $ a=\pm u/2 $ and obviously, 
Eqs. (\ref{CPA01-04}) fails for $ u>0 $.
Therefore, for all $ u>0 $ and $ \alpha $, 
$ \rho(\alpha) > 0 $ is equivalent to $ b \neq 0 $.
If all the limiting values $ a $, $ b $, and $ \xi $ such that $ b \neq 0 $ exist, 
$ G_R $ and $ G_I $ are also bounded and 
there are no indeterminate forms in Eqs. (\ref{CPA01-03}) 
in the case of $ \varepsilon \to +0 $.
In this situation, ($ a,b,\xi $) satisfies 
\begin{equation}\label{CPA01-05}
G_Ra^2+\left[1-2bG_I\right]a
-\left(b^2+\frac{u^2}{4}\right)G_R=0, \ \
G_Ia^2+2bG_Ra+b-\left(b^2+\frac{u^2}{4}\right)G_I=0.
\end{equation}
The Green function is 
\begin{equation}\label{CPA01-06}
G_R=\int^1_{-1} \frac{(\xi-x)\rho_0(x)}
{(x-\xi)^2+b^2}\text{d}x, \ \
G_I=\int^1_{-1} \frac{b\rho_0(x)}
{(x-\xi)^2+b^2}\text{d}x.
\end{equation}
We multiply $ G_I $ to the first of Eqs. (\ref{CPA01-05}) and $ G_R $ to the second of Eqs. (\ref{CPA01-05}). 
Subtracting the former from the latter, 
we obtain $ (2bG_R^2+2bG_I^2-G_I)a=-bG_R $. 
Therefore, if ($ a,b,\xi $) is the solution, it must satisfy 
either of below two sets of equations.
The first set is given by 
\begin{equation}\label{CPA01-07}
2bG_R^2+2bG_I^2-G_I \neq 0, \ \
a=\frac{-bG_R}{2bG_R^2+2bG_I^2-G_I}, \ \
u^2=\frac{-4b(bG_R^2+bG_I^2-G_I)\left[(1-2bG_I)^2+4b^2G_R^2\right]}
{\left[2bG_R^2+2bG_I^2-G_I\right]^2}. 
\end{equation}
The second set is given by 
\begin{equation}\label{CPA01-08}
G_R = 0, \ \
1-2bG_I = 0, \ \
a^2+b^2=\frac{u^2}{4}.
\end{equation}
These two sets are necessary conditions for ($ a,b,\xi $) is the solution 
of Eqs. (\ref{CPA01-05}). And obviously, 
if ($ a,b,\xi $) satisfies either Eqs. (\ref{CPA01-07}) or Eqs. (\ref{CPA01-08}), 
it is the solution of Eqs. (\ref{CPA01-05}). It means, these two sets are 
necessary and sufficient conditions for ($ a,b,\xi $) is the solution. 
Therefore, in order to constitute the energy band, 
we solve Eqs. (\ref{CPA01-07}) and Eqs. (\ref{CPA01-08}), instead of Eqs. (\ref{CPA01-05}).

In the next, we consider the MIT in the case of $ \rho_0(x)=\rho_0(-x) $. 
We define $ u_c $ as a critical value of $ u $ when the single energy band split into two regions. 
Namely, $ u < u_c $ represents a metallic state (the single band) and 
$ u > u_c $ represents an insulating state (the divided bands). 
We derive the expression of $ u_c $ analytically. 

To clarify our process of deciding $ u_c $, 
we determine the energy $ \alpha $ that the splitting of the single band 
takes place. Because $ \rho_0(x) $ is even function, 
$ G_R(b,\xi)=-G_R(b,-\xi) $ and $ G_I(b,\xi)=G_I(b,-\xi) $ are satisfied. 
As is obvious from Eqs. (\ref{CPA01-07}) and (\ref{CPA01-08}), if ($ a,b,\xi $) is the solution, 
($ -a,b,-\xi $) is also the solution for fixed $ u $. Therefore, 
if $ \rho(\alpha) > 0 $ for $ \alpha=a+\xi $, 
$ \rho(-\alpha) = \rho(\alpha) > 0 $ is also valid for $ -\alpha=-a-\xi $.
This shows the energy band is symmetric with respect to $ \alpha=0 $. 
It makes a point that the band splitting takes place at $ \alpha=0 $.
Thus, all we have to do is to find $ u $ that brings $ \rho(\alpha=0) > 0 $.
Its supreme value is $ u_c $.
We search for all the possible $ u $ 
in Eqs. (\ref{CPA01-07}) and Eqs. (\ref{CPA01-08}).

At first, Eqs. (\ref{CPA01-08}) fails for $ \xi \neq 0 $.
This is because for $ \xi \neq 0 $ and $ G_R=0 $, we can calculate as
\begin{equation}\label{CPA01-09}
1-2bG_I=
2\int_0^1\frac{(x^2-b^2-\xi^2)^2}{(x-\xi)^2+b^2}
\frac{\rho_0(x)}{(x+\xi)^2+b^2}\text{d}x>0.
\end{equation}
It means, Eqs. (\ref{CPA01-08}) can be satisfied only if $ \xi=0 $.
For $ \xi=0 $, $ G_R=0 $ is automatically valid because $ \rho_0(x) $ is even function.
In this case, there is an unique $ b < 0 $ that satisfies $ 1-2bG_I(b,\xi=0)=0 $ and we can obtain 
$ \alpha=a=\pm \sqrt{u^2/4-b^2} $.
For $ \alpha=0 $, $ u=2|b| $ should be satisfied.
As we show shortly after, this solution can be consolidated in the solution of Eqs. (\ref{CPA01-07}).
Now, we turn on Eqs. (\ref{CPA01-07}).
We rewrite $ \alpha=a+\xi $ through the second of Eqs. (\ref{CPA01-07}) and obtain
\begin{equation}\label{CPA01-10}
\alpha=\frac{-4\xi b(\xi^2+b^2)}{2bG_R^2+2bG_I^2-G_I}\times
\int_0^1 \frac{(x^2-y^2)^2\rho_0(x)\rho_0(y)\text{d}x\text{d}y}
{[(x - \xi)^2+b^2][(x + \xi)^2+b^2]
[(y - \xi)^2+b^2][(y + \xi)^2+b^2]}.
\end{equation}
It shows $ \alpha=0 $ is equivalent to $ \xi=0 $. 
With the fact that $ \xi=0 $ brings $ G_R=0 $ and $ 1-2bG_I \neq 0 $, 
the last of Eqs. (\ref{CPA01-07}) becomes 
\begin{equation}\label{CPA01-11}
u=\sqrt{4\left(\frac{b}{G_I(b,\xi=0)}-b^2\right)}.
\end{equation}
Considering $ G_I(b,\xi=0)=1/2b $ in Eq. (\ref{CPA01-11}), $ u=2|b| $ is obtained. This is the solution 
we derived from Eqs. (\ref{CPA01-08}).
If $ b $ satisfies Eq. (\ref{CPA01-11}) for a given $ u $, 
$ (a,b,\xi)=(0,b,0) $ is the solution brings $ \rho(\alpha=0) > 0 $. 
On the contrary, for all $ b < 0 $, $ \rho(\alpha=0) > 0 $ is obtained 
at the value of $ u $ which is given by the right side of Eq. (\ref{CPA01-11}). 
Therefore, we can formalize $ u_c $ as 
\begin{equation}\label{CPA01-12}
u_c=\sup_{b<0}\sqrt{4\left(
\frac{b}
{
\int_{-1}^1\frac{b\rho_0(x)\text{d}x}{x^2+b^2}
}
-b^2\right)}.
\end{equation}
This right side is monotonically decreasing function for $ b < 0 $. 
Since $ b \to -0 $ brings $ u \to +0 $, $ \rho(\alpha=0) > 0 $ is obtained 
for $ u $ that increases continuously from 0. 
Consequently, taking $ b \to -\infty $, $ u_c $ is given by 
\begin{equation}\label{CPA01-13}
u_c=\sqrt{8\int_{0}^1x^2\rho_0(x)\text{d}x}.
\end{equation}

In conclusion, we have derived the Mott MIT condition with the DOS through the equation,
\begin{equation}\label{CPA01-14}
\frac{U}{W}=\sqrt{2\int_0^1x^2\rho_0(x)\text{d}x}. 
\end{equation}
This equation is obtained under paramagnetic state, zero temperature and the DOS is even function in the framework of CPA.
We have proven the dependence of the MIT condition on the DOS concisely and quantitatively. 
Equation (\ref{CPA01-14}) asserts that, if the integral becomes larger, the condition is satisfied with smaller $ W $. 
The weight $ x^2 $ makes a larger contribution to the integral at $ |x| \lesssim 1 $ than $ |x| \simeq 0 $. 
It indicates a material metalizes easily if its eigenstates are collected near the band edge 
rather than around the band center. 
Also, the integral of Eq. (\ref{CPA01-14}) is not more than 1. 
So in any cases, $ U \le W $ is a necessary condition for the MIT.
Through this proof, particularly in the key Eqs. (\ref{CPA01-09}) and (\ref{CPA01-10}), 
all the properties of $ \rho_0(x) $ we have used are just $ \rho_0(x)=\rho_0(-x) $ and $ \rho_0(x) \ge 0 $. 
Thus, to execute Eq. (\ref{CPA01-14}), we can use the DOS $ \rho_0(x) $ as its original form (sum of delta functions \cite{1996Blawid}) 
and there is no need to approximate the DOS by a continuous function \cite{1963Hubbard, 1973Yonezawa1, 1983Ishikawa}. 
Once all the eigenvalues for single electron is obtained, $ \rho_0(x) $ is composed and 
the MIT condition can be calculated analytically.

\begin{acknowledgments}
We thank T. Ogawa and T. Ohashi for useful discussions.
\end{acknowledgments}


\end{document}